\documentstyle[aps,twocolumn,graphicx]{revtex}

\begin{document}

\title{
Spontaneous structure formation in a network of chaotic units\\
with variable connection strengths
}

\author{
Junji Ito and Kunihiko Kaneko
}

\address{
Department of Pure and Applied Sciences,\\
College of Arts and Sciences, University of Tokyo\\
3-8-1 Komaba, Meguro-ku, Tokyo, 153-8902 Japan\\
}

\date{\today}

\maketitle

\begin{abstract}
As a model of temporally evolving networks, we consider a globally
coupled logistic map with variable connection weights. The model 
exhibits self-organization of network structure, reflected by
the collective behavior of units. Structural order emerges even without 
any inter-unit synchronization of dynamics. Within this structure, units 
spontaneously separate into two groups whose distinguishing feature is 
that the first group possesses many outwardly-directed connections to the 
second group, while the second group possesses only few outwardly-directed 
connections to the first. The relevance of the results to structure 
formation in neural 
networks is briefly discussed.
\end{abstract}


Recently, studies of various types of networks have been
attracting the interest of researchers from a broad range of scientific
diciplines \cite{small-world,scale-free}. 
Although in some such studies, 
there is a time dependence of the structure of the network consisting
of an increase in the number of units of which it is composed 
\cite{scale-free}, the
elements themselves are essentially static units, 
with neither their intrinsic properties nor their interactions with
each other evolving in time.
Most networks in the real world, however, consist
of dynamic elements, and the dynamics of the individual elements
influence the formation of network structure.
Thus in order to understand realistic networks, it is necessary to 
consider the modeling of networks with such dynamic elements.
In this Letter we study an abstract model of a network composed of 
dynamic elements and report its behavior, as found through numerical 
simulation, focusing mainly on the formation of structure.


We consider a network of $N$ dynamic units that interact
with each other through connections with time-dependent strengths.
For simplicity,
we describe the dynamics of both the units and the connection
strengths with discrete-time maps. Hence, our model belongs to a
class of globally coupled maps (GCM) \cite{GCM}.  We denote by $f(x)$
the function defining the map for the dynamics of each unit. With this,
our model is given by the set of equations
\begin{equation}
\label{model_unit.eq}
 x^{i}_{n+1} = f[(1-c) x^{i}_{n}+c \sum^{N}_{j=1} w^{ij}_{n}x^{j}_{n}],
\end{equation}
where $x^{i}_{n}$ is the state variable of the $i$-th unit ($1\leq i
\leq N$) at the $n$-th time step.  The coupling $c$  represents
the strength of the influence of the other units on the dynamics
of unit $i$ ($0<c<1$), and $w^{ij}_{n}$ is the time-dependent
weight of the connection from the unit $j$ to $i$ at time step $n$.
As the map providing the dynamics of the units, we adopt the logistic map
$f(x)=ax(1-x)$, but we believe that qualitatively similar behavior would 
be displayed by the system for any form of $f(x)$ that exhibits chaos.

With regard to the dynamics of the connection strengths, we stipulate
that the connections between units $i$ and $j$ with similar values
$x^{i}_{n}$ and $x^{j}_{n}$ are strenghened \cite{Rel,Ito}. 
(This can be regarded
as an extension of Hebb's rule, which is widely used
in neural network studies \cite{neuralnet}.) Also, we consider there to be
a resource in the system that is used to establish connections between units.
Then, we assume that there is a limitation on this resource.
As a result, there exists competetion among connections 
for this resource. Instead of using an explicit variable representing 
the resource, we incorporate this effect into our model through the
normalization of the connection strengths, as
\begin{equation}
\label{model_connection.eq} w^{ij}_{n+1} = \frac{[1 + \delta \cdot
g(x^{i}_n, x^{j}_n)] w^{ij}_n}{\sum_{j=1}^{N} [1 + \delta \cdot
g(x^{i}_n, x^{j}_n)] w^{ij}_n},
\end{equation}
where $\delta$ is a parameter that represents the plasticity of
the connection strengths, and $g(x^{i}_n, x^{j}_n)$ is a
monotonically decreasing function of the absolute value of the
difference between its
arguments, whose form we choose here is $g(x^{i}_n, x^{j}_n)= 
1-2|x^{i}_n - x^{j}_n |$.
Note that, due to the normalization given in Eq.
(\ref{model_connection.eq}), $w^{ij}$ is generally not equal to
$w^{ji}$; i.e., the network is asymmetric.

In the following, we give the results of our numerical simulations
of the model. Throughout this Letter, the number of units $N$
is set to $100$, though the results described below do not
change qualitatively for larger systems except for 
the existence of a longer transient behavior. The
initial conditions we used are as follows. First, the initial
values of the self-connections $w^{ii}_{0}$ were set to $0$.
Then as determined by Eq. (\ref{model_connection.eq}), they remained at $0$ for $n>0$.
Second, all the remaining connection strengths were set to
identical values. From the constraint of the normalization, this value is
determined to be $1/(N-1)$. Finally, for the state variables, the
initial values were randomly chosen from the interval $(0,1)$ with a uniform sampling measure.


In our model, we have three parameters: $a$, which controls the dynamics of each unit, $c$, which determines the overall
sterngth of the interactions
between the units, and $\delta$, which governs the connection
dynamics. Here we fix the parameter $\delta$ to $0.1$ \cite{com1}
and study how the behavior of the system changes as the function of the
values of the parameters $a$ and $c$.

It is known that the dynamics of GCM can be classified into 
four phases, according to the degree of synchronization and 
clustering among units \cite{GCM}.
In contrast to the conventional GCM, only three of these four
phases appear in our system. The first is the coherent phase,
in which all the units take the same value and oscillate
synchronously. The second is the ordered phase, in which the units
split into a few clusters and all the units within each such cluster 
oscillate synchronously. The third is the desynchronized phase, in which there
is no synchronization between any two units \cite{com2}.

Corresponding to the different types of collective behavior, different
network structures emerge. In the coherent phase, all connections
have almost identical values and these values do not change over
time. In the ordered phase, due to the formation of
clusters of units, connections between units that belong to
the same cluster have similar finite values, determined by the
size of the cluster, while connections between two units from 
different clusters tend towards $0$. In this case too, the network is static. 
The situation is different, however, in the desynchronized phase,
in which connection strengths can change, and the network structure
is not fixed over time. The structure of the network in this phase
is complicated, but not completely random. In this Letter we
consider only the phenomena observed in this desynchronized phase,
because we are presently interested in the behavior of dynamic networks. This
phase roughly corresponds to the parameter ranges $3.7<a<4.0$ and
$0<c<0.2$. In the simulations reported in th following, parameter values outside these ranges
were not used.

To characterize the global behavior of the network in the
parameter space, we define some characteristic quantities of the
network and study their parameter dependence.

First, as an index of the magnitude of the temporal change of
the network, we define an average variation of the connection
strength per step. We call this the `activity of the network',
and write it
\begin{equation}
\label{activity.eq} A = \frac{1}{(N-1)^2} \cdot
\frac{1}{\tau_{m}}\sum_{i \ne j}
\sum_{n=\tau_{t}}^{\tau_{t}+\tau_{m}} | w_{n}^{ij} - w_{n-1}^{ij}
|,
\end{equation}
where $\tau_{t}$ is the length of the transient period and 
$\tau_{m}$ is the length of the measuring period.

In FIG. 1(a), the activity $A$ of the network is plotted
with respect to the parameters $a$ and $c$ on a gray scale. Here
$\tau_{t}$ and $\tau_{m}$ were chosen as $100,000$ and $1,000$, 
respectively. A broad band of high activity is seen around the 
line $c = .15 \times (a-3.7)$, corresponding to the bright region in FIG. 1(a).
Note that there is no synchronization between the dynamics of the units
anywhere in the parameter space shown in Fig 1(a), 
 Nevertheless, 
there is a rather wide region of quite low activity. In most of this region, 
most of th units exist in pairs, with the units in each such pair having 
non-zero connections only between each other, 
forming fixed pairs in the network. While the dynamics of two units forming 
a pair are not synchronized, they are highly correlated.

In the regime of high activity, more complex and dynamic network
structure is formed. To observe this, we consider the average
connection matrix, denoted as $W^{ij}$ and defined as the temporal
average of $w_{n}^{ij}$: $W^{ij} = \frac{1}{\tau_{m}}
\sum_{n=\tau_{t}}^{\tau_{t}+\tau_{m}} w_{n}^{ij},$ where
$\tau_{m}$ and $\tau_{t}$ were introduced in 
Eq. (\ref{activity.eq}). 

If the dynamics of the connection strength are completely random, 
it is expected that the average connection strengths will take almost 
identical values for each $i$ and $j$, and the variance among units will
decrease to 0 as the averaging 
time increased. Contrastingly, if there exists structure in a
network with high activity, there should be some variance among 
units in $W^{ij}$. Keeping this in mind, we consider the sum of 
the average connection strengths eminating from one unit: $W_{out}^{i} 
= \sum_{j=1}^{N} W^{ji}$. We calculated the variance of $W_{out}^{i}$ 
over $i$ for different parameter values. The result is displayed in 
FIG. 1(b) for $\tau_{t}=100,000$ and $\tau_{m}=1,000$. 
We find that a large variance is observed just below the line $c = .15 \times(a-3.7)$.

Comparing FIG.s 1(a) and (b), we can see that the region of high
network activity can be decomposed into two regimes: one with 
a large variance of $W_{out}^{i}$ 
[for $c < .15 \times(a-3.7)$], 
and the other with small $W_{out}^{i}$ 
[for $c > .15 \times(a-3.7)$]. 
As mentioned above, a large variance in the active 
regime indicates the existence of some structural order in a temporally 
evolving network.

In the following, we investigate the structural characteristics of
this dynamic yet ordered network. In particular we consider the parameter
values $a=3.97$ and $c=0.12$, which correspond to the largest
variance in the active regime However we point out that the general 
characteristics of
the network do not depend sensitively on this special choice of 
the parameter values.

First, we study the structural change over time from
the initial all-to-all type network to the eventual highly
structured one, by considering the dependence on the averaging time
of  $W_{out}^{i}$. In FIG. 2, 
we plot series of $W_{out}^{i}$ as functions of the measuring time
$\tau_{m}$ (with fixed transient time $\tau_{t}=0$) for a single
trial \cite{com3}. Each line represents a series of $W_{out}^{i}$ 
for a particular value of $i$. 

This figure shows that units separate into two
groups: one with large values of $W_{out}^{i}$ and one with small values
\cite{com4}.The separation becomes more distinct 
as the measuring time increases, 
although the separation process seems to be nearly completed 
by the $3 \times 10^{6}$-th step, 
because after this time, we do not observe the migration of any unit 
between the two groups.
Also, this figure shows that
the fluctuations of $W_{out}^{i}$ are larger for the large $W_{out}^{i}$
group. This implies that $W_{out}^{i}$ for a unit in this
group occasionally takes small values for a certain period. 
By contrast, a unit of the small $W_{out}^{i}$ group will only very rarely 
take large values of the total 
weight. In this sense, the small $W_{out}^{i}$ group is more stable.

To quantify the detailed properties of the network structure, we
digitize the connections as follows: If $w^{ij}$
exceeds a threshold value, namely $1/(N-1)$, we assign a
connection from unit $j$ to unit $i$; otherwise no connection is
assigned. This threshold $1/(N-1)$ is equal to
the connection value in the case that
a unit uniformly connects to all the others. Hence it is a natural
criterion for distinguishing `strong' connections.

Using this method, we can represent the network by a graph. We
composed graphs from snapshots of $w^{ij}$ at the $500,000$-th step for
many initial conditions. From these graphs, we calculated
distribution of the degree of reception and emission.
The degree of reception is the number of connections directed at
a unit (subsequently reffered to as ``inwardly-directed'' connections)
and the degree of emission is the number of connections 
eminating from a unit (subsequently reffered to as ``outwardly-directed
connections''.) The distributions of these two quantities
are shown in FIG. 3.

The distribution of the degree of reception has a unimodal shape,
with a peak at about $8$ degrees. Hence, with regard to the inwardly-directed
connections, this network has a single scale. This is mainly due
to the competition among inwardly-directed connections, resulting from their
normalization.

By contrast, the distribution of the degree of emission has
a bimodal shape, which can be decomposed into two components. One 
component is 
a distribution with exponential decay, corresponding to the
small $W_{out}^{i}$ group in FIG. 2. The nother component
is the unimodal distribution with
a peak at about $45$ degrees, corresponding to the large $W_{out}^{i}$ 
group in FIG. 2. 

Considering the appearance of the two components in the
distribution of the outwardly-directed connections, we divide the units into
two groups. One group we call the `core group,' which consists of units
with more than $20$ connections, and the other we call the
`peripheral group,' which consists of all other units.
With the parameter values used here, the number of units in the
core group is typically 14.

With the partition of units into these two groups, the
connections are naturally classified into four groups. 
The group to which a given connection belongs is determined by the
groups to which the two units it connects belong.
With obvious identification, we call these groups the `core-to-core
group', the `core-to-peripheral group', the 'peripheral-to-core group',
and the `peripheral-to-peripheral group'. 

The number and density of connections in each group are listed
in Table 1. The most apparent characteristics seen here are that the
peripheral-to-core group has very few members, while the
core-to-peripheral group has many members. 
Also, it is senn that the density of core-to-core connections is quite
high and that of peripheral-to-peripheral connection is quite low.
From these results, we can conclude 
that the units in the core group interact strongly with each other and
that the dynamics of the core group
strongly influence the peripheral group, but that the dynamics of the
peripheral group have almost no influence on the core group.

To this point, we have investigated our system mainly with regard to
network structure, largely ignoring the dynamics of units
underlying the structure formation. 
However, it is clear that there must be
interdependency of the unit dynamics and connection dynamics
for the structure formation discussed above to occur. 
We have confirmed this directly through numerical simulations.
Using the somewhat unnatural restriction under which connection weights 
depend on the dynamics of the units but that the dynamics of the units
do not depend on the connection weights, we found that no structure ever
appears.
More detailed analysis of this point is now underway and
its results will be presented elsewhere.


To summarize, as a model of temporally evolving networks, we
have considered a network of dynamic units, whose dynamics are described
by logistic maps and which are coupled to each other with 
variable connection weights.
The model exhibits dynamical self-organization of its network
structure, reflected by the state of units' collective behavior.
Even in the parameter region where there is no synchronization of
the unit dynamics, some structural order emerges. There, units
spontaneously separate into two groups, with one group possessing
especially many outwardly-directed connections to the other group. 

Because of the simplicity of the model and the universality among globally 
coupled maps, we believe that 
the phenomena revealed in this study are exhibited generally by
a network whose connections change in a manner governed by
the relationships between its dynamic elements.
One example is neural networks. Though the conventional understanding
has been that the timescale of the change of synaptic weights is
much slower than that of neuronal dynamics, more and more
evidence is being published providing evidence that synaptic change
occurs over a wide range of timescales, from hundreds of milliseconds to
months or years \cite{fast-synapse}. In addition, it is known 
that, in the early stage of development, axons arborize excessively, 
and eventually are trimmed under the influence of neuronal activity
\cite{arborize}. Our model seems to be suited to the modeling of 
such a situation. The structure formation observed in this study
may provide a basic description of
local structure formation in the brain, such as columnar structure.

As mentioned at the beginning of this Letter, there is yet little known about
dynamic networks. The significance of our study will be revealed
as empirical data about dynamic networks are obtained.

The authors would like to thank T. Ohira for his valuable suggestions,
and to G. C. Paquette for the critical reading of the manuscript.
This work is supported by Grants-in-Aid for Scientific Research 
from Ministry of Education, Science and Culture of Japan (11CE2006).

\begin{figure}[t]
\begin{center}
\includegraphics[width=8cm]{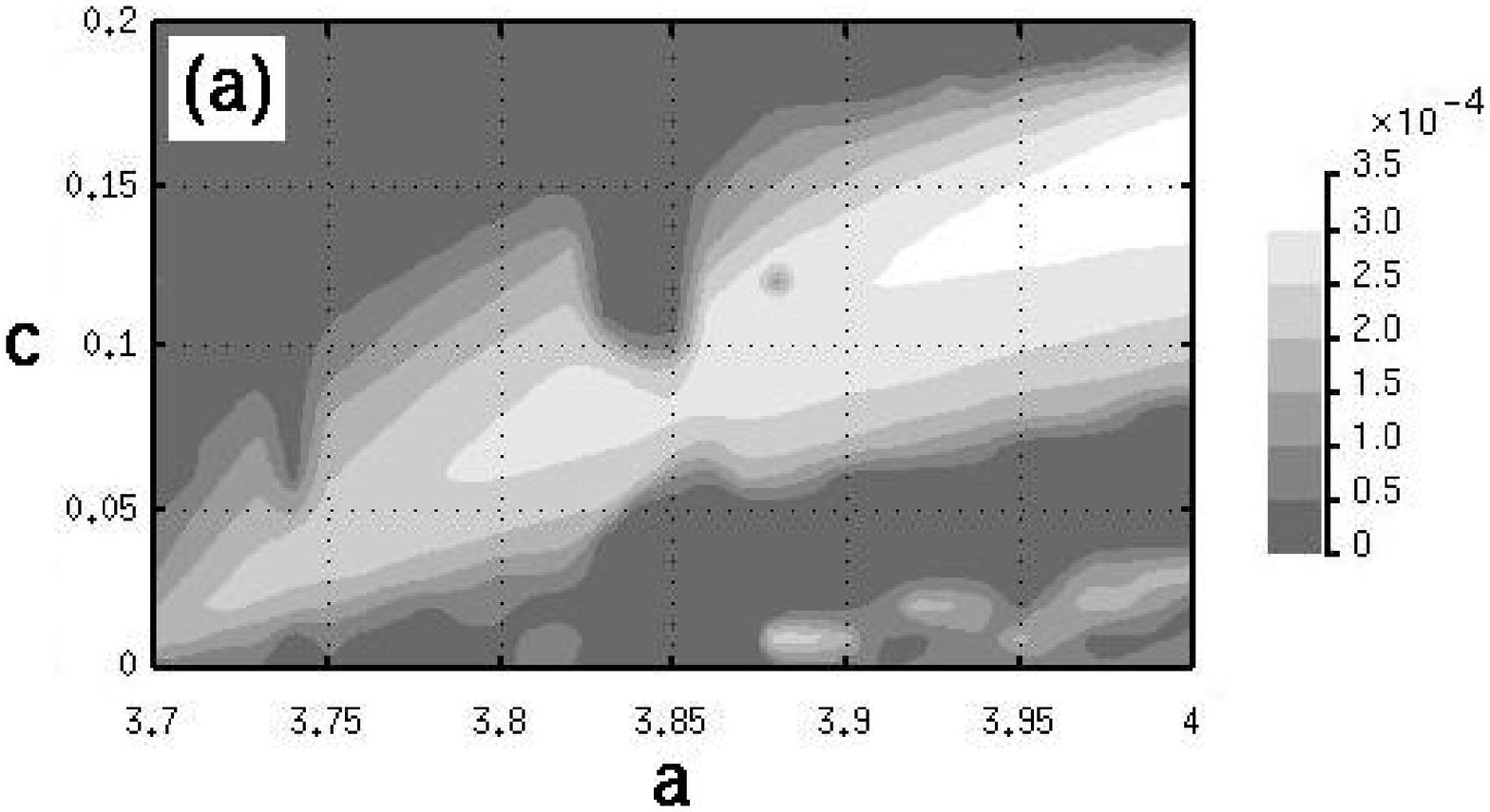}
\includegraphics[width=8cm]{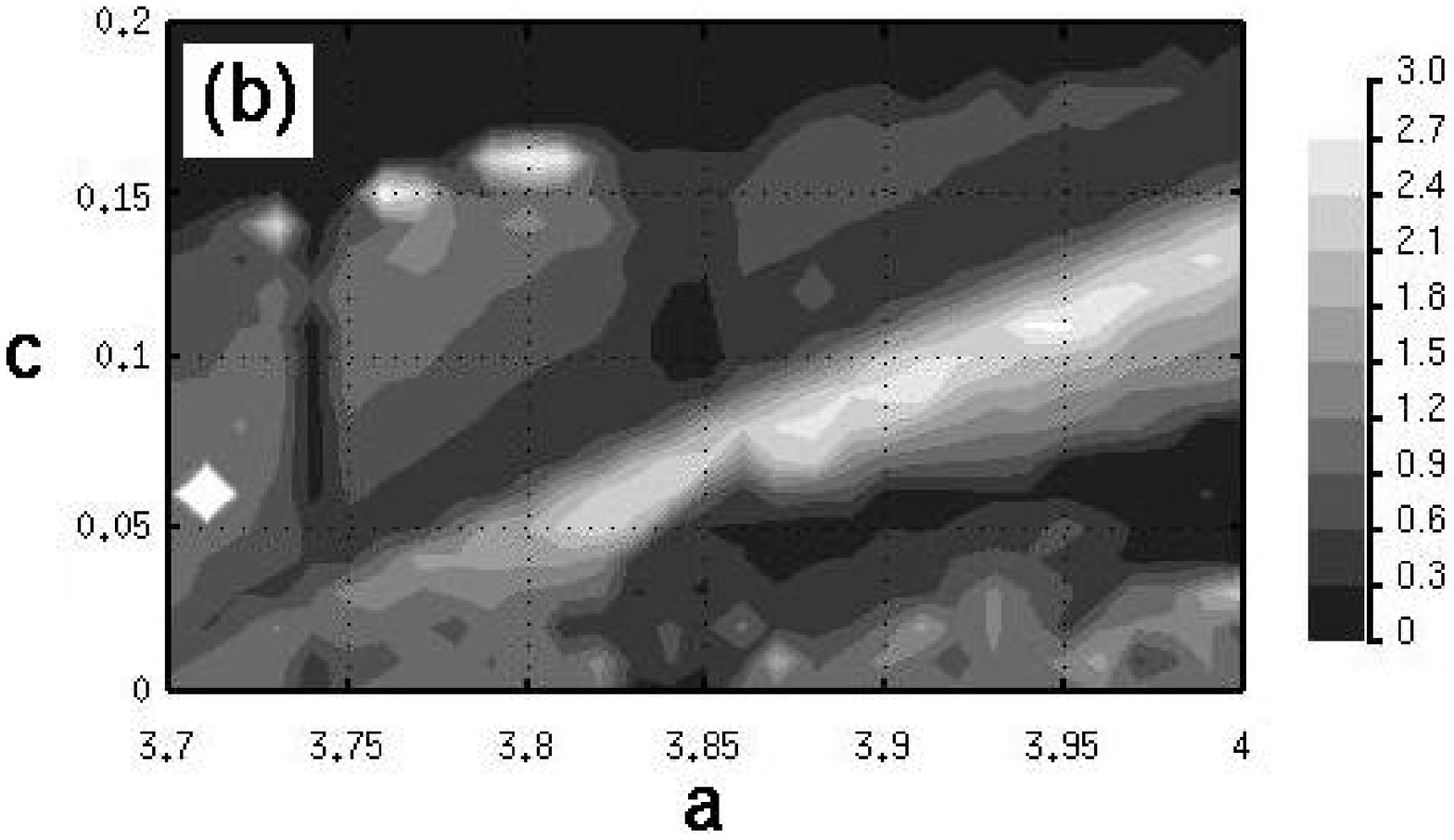}
\end{center}
\caption{ Gray scale plots of network quantities with respect to the
parameters $a$ and $c$, with a discretization of 0.01 for both.
(a) The activity $A$ of the network.
Brighter color corresponds to higher network activity. (See the
text for the definition of the quantity $A$.) (b) The variance of
the total weight of outwardly-directed connections $W_{out}^{i}$. Brighter
color corresponds to larger variance.}
\end{figure}

\vspace{-0.5cm}
\begin{figure}[t]
\begin{center}
\includegraphics[width=8cm]{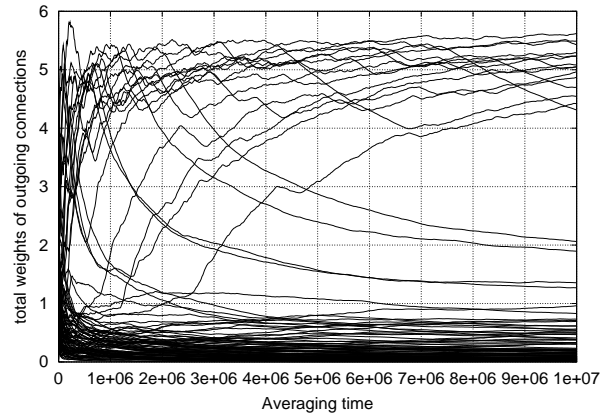}
\end{center}
\caption{The total weights of outwardly-directed connections
$W_{out}^{i}$ as functions of the measuring time for
the average connection matrix $W^{ij}$. Each curve 
corresponds to a single unit and represents a series
of $W_{out}^{i}$ with different measuring times.
The series of $W_{out}^{i}$ for all units are superimposed.}
\end{figure}

\vspace{-0.5cm}
\begin{figure}
\begin{center}
\includegraphics[width=8cm]{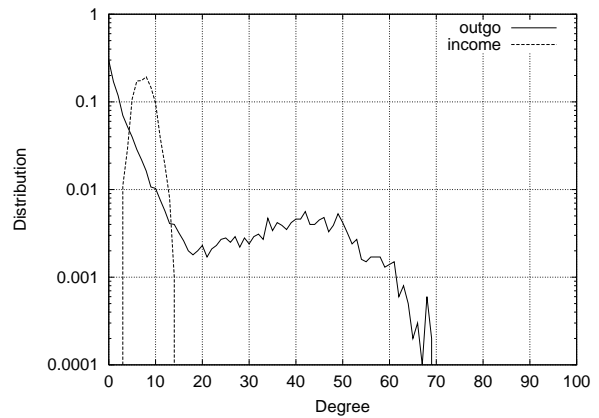}
\end{center}
\caption{ Distributions of the degrees of reception and emission.
A logarithmic scale is used for the vertical axis. The solid curve
is the distribution of the degree of emission, and the broken line 
is that of the degree of reception. }
\end{figure}

\vspace{-0.5cm}
\begin{table}
\begin{tabular}{ccr}
Group & No. of connections & density \\
\hline
CC & 159 & 0.6625 \\
CP & 792 & 0.5893 \\
PC & 2 & 0.0015 \\
PP & 234 & 0.0336 \\
\end{tabular}
\caption{Number and density of connections in each group. 
Here, CC, CP, PC and PP denote the core-to-core, core-to-peripheral,
perhipheral-to-core and peripheral-to-peripheral groups.
(See the text for the definitions of the groups.)}
\end{table}


\begin{thebibliography}{99}
\bibitem{com1}
Qualitatively similar behavior is observed over a wide range
of values of $\delta$, but for very small values, (say $\delta<.01$,)
we have not 
observed the network structure discussed here. The timescales 
of the unit dynamics and connection dynamics must be set to be of 
the same order.
\bibitem{com3}
Here we have not plotted the timeseries of $W_{out}^{i}$ themselves,
as we wish to illustrate the separation of the units clearly.
\bibitem{com4}
In FIG. 2, the separation is amplified due to longer averaging 
times for larger $\tau_m$. 
However, we have confirmed from a direct plot of $W_{out}^{i}$ that 
after transient behavior dies out, if a given $W_{out}^{i}$ ever takes
a sufficiently small value, it never again joins the large $W_{out}^{i}$
group, and if it ever takes a sufficiently large value, it never again 
joins to the small $W_{out}^{i}$ group. Also, it should e noted that 
the fluctuations of the $W_{out}^{i}$ in the upper group are larger than
those in the lower group, but even with ther large fluctuations,
these $W_{out}^{i}$ do not fall to the level of the small $W_{out}^{i}$
group.
\end{thebibliography}
\end{document}